\documentclass[
reprint,
nofootinbib,
 amsmath,amssymb,
 aps,
prd,
]{revtex4-1}
\usepackage{lipsum}
\usepackage{dcolumn}
\usepackage{bm}
\usepackage{color}
\usepackage{graphicx,amsmath,amssymb}
\usepackage{tikz}
\usetikzlibrary{matrix}

\newcommand{\eqnsplit}[1]{\begin{align}\begin{split}#1\end{split}\end{align}}
\newcommand{\eqn}[1]{\begin{align}#1\end{align}}

\definecolor{awesome}{rgb}{1.0, 0.13, 0.32}
\definecolor{azure(colorwheel)}{rgb}{0.0, 0.5, 1.0}
\definecolor{scarlet}{rgb}{1.0, 0.13, 0.0}
\definecolor{springgreen}{rgb}{0.0, 1.0, 0.5}

\usepackage{hyperref}
\hypersetup{
    bookmarks=true,         
    unicode=false,          
    pdftoolbar=true,        
    pdfmenubar=true,        
    pdffitwindow=false,     
    pdfstartview={FitH},    
    pdftitle={My title},    
    pdfauthor={Author},     
    pdfsubject={Subject},   
    pdfcreator={Creator},   
    pdfproducer={Producer}, 
    pdfkeywords={keyword1, key2, key3}, 
    pdfnewwindow=true,      
    colorlinks=true,       
    linkcolor=azure(colorwheel),          
    citecolor=awesome,        
    filecolor=azure(colorwheel),      
    urlcolor=blue     
}

\begin{document}
\title{Super Gaussian enhancers in the Schwinger mechanism}

\author{Ibrahim Akal}
\email{ibrahim.akal@desy.de}
\affiliation{Theory Group, DESY\\22607 Hamburg, Germany}

\date{\today}
\begin{abstract}
We study the Schwinger mechanism in the presence of an additional uniformly oriented, weak super Gaussian of integer order $4N+2$. 
Using the worldline approach, we determine the relevant critical points to compute the leading order exponential factor analytically. 
We show that increasing the parameter $N$ gives rise to a strong dynamical enhancement. For $N=2$, this effect turns out to be larger compared to a weak contribution of Sauter type. 
For higher orders, specifically, for the rectangular barrier limit, i.e. $N \rightarrow \infty$, we approach the Lorentzian case as an upper bound.
Although the mentioned backgrounds significantly differ in Minkowski spacetime, we show that the found coincidence
applies due to identical reflection points in the Euclidean instanton plane.
In addition, we also treat the background in perturbation theory following recent ideas. By doing so, we show that the parameter $N$ determines whether
the weak contribution behaves perturbatively or nonperturbatively with respect to the field strength ratio and, hence, reveals an interesting dependence on the background shape. In particular, we show that 
for backgrounds, for which higher orders in the field strength ratio turn out to be relevant, a proposed integral condition is not fulfilled. In view of these findings, the latter may serve as an indicator for the necessity of higher order contributions.
\end{abstract}
\maketitle
\section{Introduction}
Tunnelling of matter-antimatter pairs from the quantum vacuum in a background gauge field 
is an important nonperturbative prediction in quantum field theory
\cite{Schwinger:1951nm}.
For charged particles\footnote{The charge has been absorbed into the field strength. Throughout this paper we use natural units $c = 1$ and $\hbar = 1$.} with mass $m$ the rate in the weakly coupled regime is exponentially suppressed below the critical field strength
$E_\mathrm{S} = m^2$.
Due to the extremely large value, this so-called Schwinger mechanism still could not yet be seen in the laboratory. 

Recently, there has been made progress in investigating this mechanism in analogous condensed matter systems 
revealing interesting similarities between nonlinear quantum vacuum phenomena and nonequilibrium condensed matter systems
\cite{PhysRevLett.95.137601,Allor:2007ei,Katsnelson:2012cz,Zubkov:2012ht,Fillion-Gourdeau:2015dga,Linder:2015fba,Akal:2016stu,Fillion-Gourdeau:2016izx,Akal:2018txb,Taya:2018eng}.

Temporal inhomogeneities can trigger an enormous enhancement of the tunnelling rate \cite{Popov:1971iga,PhysRevD.2.1191,Dunne:2005sx}.
For instance, one may consider
a background composed of a strong, locally static part superimposed with an additional weak but rapid alteration \cite{Schutzhold:2008pz,Akal:2014eua}. 
Such composite backgrounds give rise to 
certain critical points \cite{Schutzhold:2008pz,Linder:2015vta,Akal:2017ilh} which act as reflectors in the instanton plane resulting in a drastic dynamical enhancement. 

Generally, the microscopic details of the weak dependence can be very decisive.
However, even alterations with a substantially distinct analytic structure in Minkowski spacetime can lead to the same rate if
the associated critical points in the Euclidean instanton plane perfectly coincide \cite{Akal:2017ilh}.

Recently, such a coincidence for the leading order exponential factor has been observed
between a weak Lorentzian and a super Gaussian approaching the rectangular barrier limit \cite{Akal:2017ilh,Akal:2017sbs}.

In this work, we go beyond previous observations and study the mentioned enhancement effects fully analytically for weak super Gaussians of general order $4N+2$ where $N \in \mathbb{N}_{> 1}$.
The reason why we consider this particular class of backgrounds is, because they allow a transition from perturbative to nonperturbative dependence on the weak field strength which is controlled by the parameter $N$.
In this way, we can directly obtain useful insights into the role of the explicit background shape.
To be more precise, we here investigate two different aspects. 

Sec.~\ref{sec:nonperturb} deals with the described coincidence for two substantially distinct backgrounds. In order to do so, we 
work within the worldline formalism in quantum field theory \cite{Strassler:1992zr,Schubert:2001he}
and utilize the reflection approach to study the behavior for different $N$. 
In particular, we derive an approximate formula for arbitrary $N$ which has not been performed so far.
We compare our findings with well studied background shapes. 
Note that numerical studies based on an alternative kinetic approach for the standard, nonassisted dynamical Schwinger process have previously been discussed only for super Gaussian pulses of moderate order \cite{Abdukerim:2013vsa}.

In Sec.~\ref{sec:perturb}
we treat the weak contribution in perturbation theory by following recent ideas put forward in \cite{Torgrimsson:2017pzs}. We explore the role of higher orders in the field strength ratio parameter. This approach usually requires the computation of Fourier transforms of the respective backgrounds which, generally, can be highly difficult to obtain. Recent studies have therefore focused on well known cases in which the transforms are relatively easy to compute, see \cite{Torgrimsson:2017pzs}. 

In the present work, we introduce the convolution technique which substantially simplifies the problem at hand and allows to derive purely analytical results for weak pulses of super Gaussian type. In particular, we find that in situations where higher orders in the field strength ratio turn out to be relevant, a proposed integral condition is not fulfilled. In view of these observations, the latter may be seen as an indicator for the relevance of higher order contributions.
%

\section{Nonperturbative approach}
\label{sec:nonperturb}
\begin{center}
\begin{figure}[h]
  \includegraphics[width=.3\textwidth]{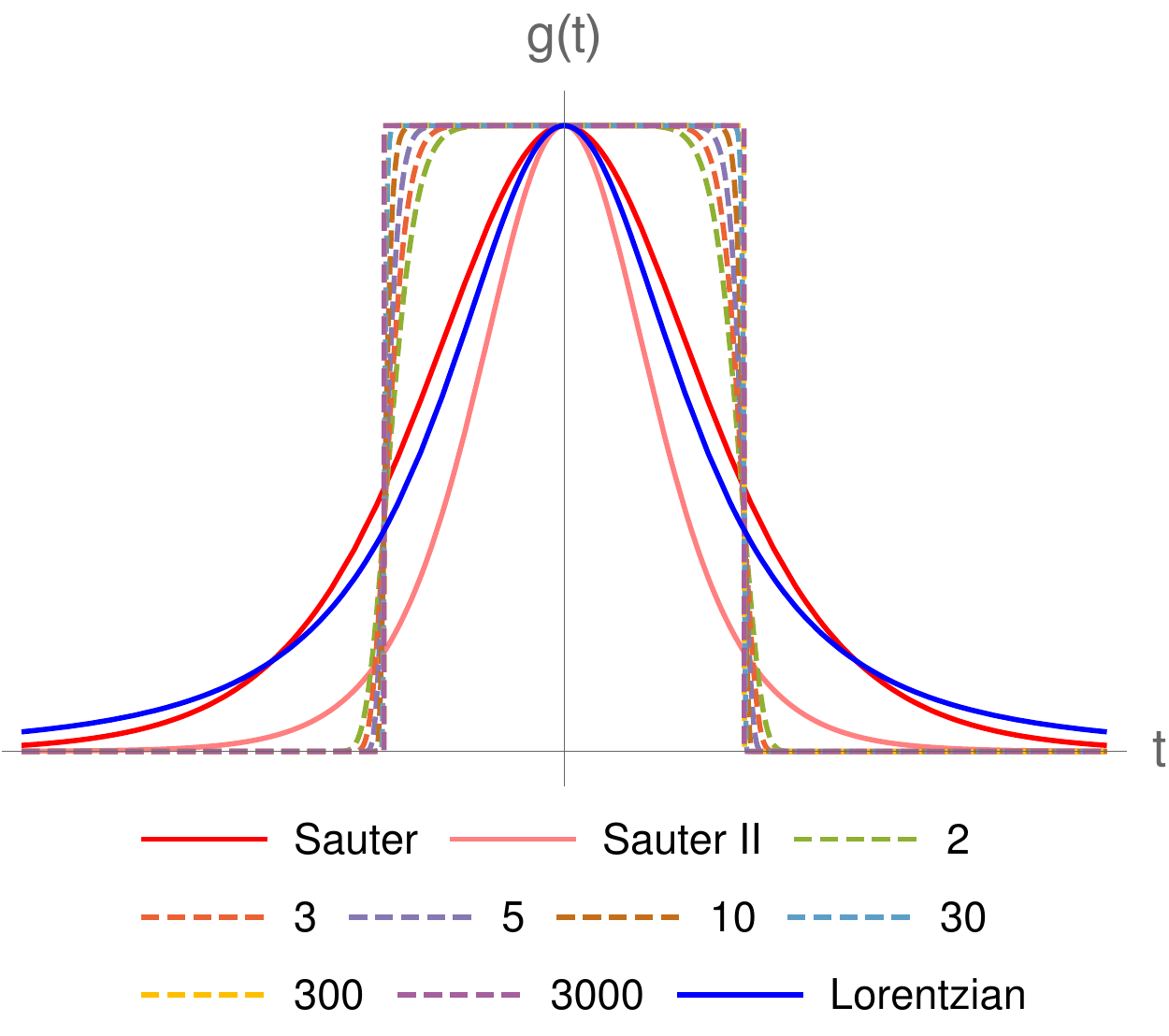}
\caption{Comparison of function $g$ plotted versus $t$. The numbers in the legend correspond to the integer $N$ in \eqref{eq:f-g-functions}. The pink curve corresponds to a modified Sauter pulse with frequency shift $\omega \rightarrow \omega \pi/2$ leading to the same $\mathcal{W}_0$ as the Lorentzian (blue). For $N \rightarrow \infty$ we approach the usual rectangular potential barrier.}
\label{fig:profiles}
\end{figure}
\end{center}

The general form for the tunnelling probability is given as
\eqnsplit{
\mathcal{P} = 1 - e^{- 2 \Gamma}
}
where the rate, $\Gamma$, is determined by the imaginary part of the Euler-Heisenberg effective action \cite{Dunne:2004nc}. 
Due to simplifications, we focus on spin zero particles.
Furthermore, we
restrict ourselves to the adiabatic, nonperturbative regime and 
neglect contributions from the dynamical gauge field.
The rate is of the form
\begin{align}
 \Gamma = \mathcal{Q} e^{-\mathcal{W}_0}.
 \label{eq:Gamma}
 \end{align}
The stationary action $\mathcal{W}_0$ in the exponent is obtained after evaluating the worldline action
\begin{align}
 \mathcal{W} = m a + i \oint du\ \dot x \cdot \mathcal{A}(x_\mu)
\end{align}
on the periodic instanton path \cite{Affleck:1981bma}
determined by
\begin{align}
 m \ddot x_\mu = i a \mathcal{F}_{\mu \nu} \dot x_\nu.
 \label{eq:instanton-eqs}
\end{align}
Since the exponential factor in $\Gamma$ is the dominant quantity \cite{Dunne:2006st,Dietrich:2007vw} for the present study,
we set the quantum fluctuation prefactor $\mathcal{Q}$ to unity.
The kinematic invariant obeys the relation $a^2 = \dot x^2$ due to the anti-symmetry of the field tensor $\mathcal{F}_{\mu \nu}$.
We consider a purely electric background which is a uniformly oriented superposition described by
\begin{align}
\pmb{E}(t) = E \left( f + \epsilon g \right) \hat x_3
\label{eq:electric-bground}
\end{align}
where $\epsilon \ll 1$ and
\begin{align}
f(t) = 1,\quad g(t) = e^{-(\omega t)^{4N +2} },\quad N \in \mathbb{N}.
\label{eq:f-g-functions}
\end{align}
In Fig.~\ref{fig:profiles} the function $g$ is depicted for various $N$ including the Sauter and Lorentzian cases.
After the rotation in the complex plane ($t \rightarrow i x_4$), we arrive at
\eqnsplit{
  \mathcal{A}_3(x_4) = 
  - i E (F
  + \epsilon G),
  \label{eq:eucl-vecpotential}
}
where 
\eqnsplit{
F(x_4) &= x_4,\\
G(x_4) &= - \frac{1}{ \omega} \frac{ ( \omega x_4) \mathbf{E}_{\frac{4N+1}{4N+2}}(-( \omega x_4)^{4N+2})}{4N+2}.
}
Here, $\mathbf{E}_n$ denotes the exponential integral function.
Inserting the vector potential \eqref{eq:eucl-vecpotential} into the instanton equations \eqref{eq:instanton-eqs}, we find the following coupled system of differential equations
\eqnsplit{
  \ddot x_4 &= + \frac{a E}{m} \big[ F^\prime + \epsilon G^\prime \big] \dot x_3,\\
  \ddot x_3 &= - \frac{a E}{m} \big[ F^\prime + \epsilon G^\prime \big] \dot x_4.
  \label{eq:full-instanton-eqs}
}
The prime denotes the derivative with respect to $x_4$.
For conventional reasons, we introduce the dimensionless combined Keldysh parameter \cite{Schutzhold:2008pz}
\eqnsplit{
  \gamma = \frac{m \omega}{E}.
}
\begin{center}
\begin{figure}[h]
  \centering
  \includegraphics[width=.48\textwidth]{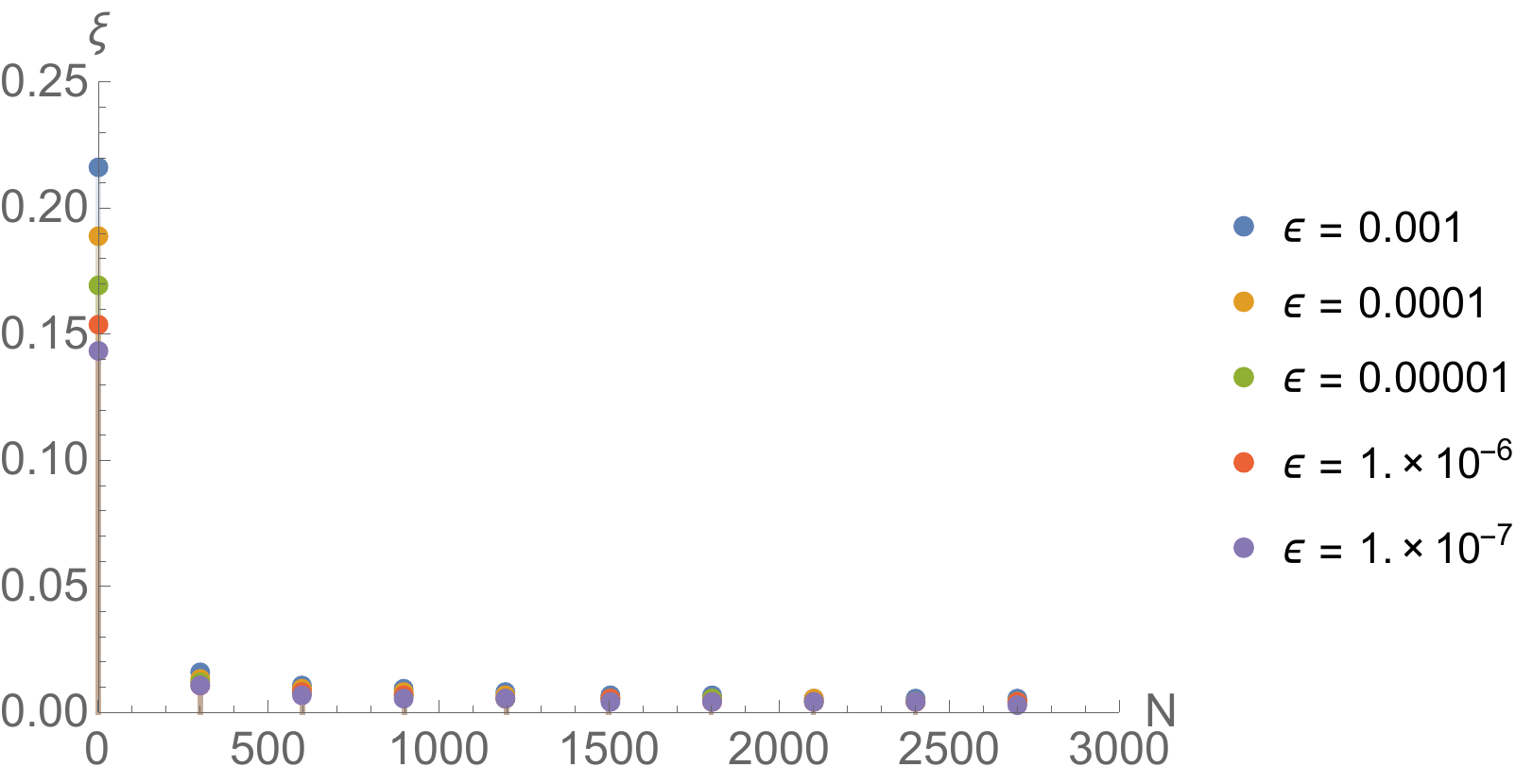}
\caption{Comparison of $\xi = - \mathfrak{Z}/(\alpha \mathfrak{D})$ versus $N$ (starting with $N=1$) for various $\epsilon$ given in the plot legend. With increasing $N$ the dependence on $\epsilon$ gets suppressed. For $N \rightarrow \infty$ we approach the Lorentzian case, i.e. $\delta = \xi \check \gamma \rightarrow 0$ (since $\xi \rightarrow 0$) and $\check \gamma \rightarrow 1$.}
\label{fig:xi}
\end{figure}
\end{center}
The idea is to compute those points for which the strong contribution can be taken as negligible compared to the weak term. 
This can be done by using an additional condition determining the complex turning points in the equivalent WKB approach. For one-dimensional temporal backgrounds, as considered here, the latter leads to the correct tunneling exponent. The relevant equation we have to solve is
\eqn{
\epsilon G(x_4^*) = \gamma/\omega.
\label{eq:rel-eqn}
}
Once $x_4^*$ has been computed it can be applied as an effective reflection point in the instanton plane.
This allows to find a sufficiently accurate expression for the stationary worldline action. For further details we would like refer to \citep{Akal:2017ilh}.

Also note that, recently, it has been discussed that further exactly solvable models may be constructed by utilizing an appropriately chosen deformation map \cite{Akal:2018hss}. This may be helpful to treat such kind of backgrounds even fully analytically. 
Interestingly, an appropriate modification of the background shape can lead to time scale reductions in
driven quantum systems, see e.g. \cite{holthaus1992pulse}.
Therefore, one may think about analogies related to such reflection points placed on the Euclidean time axis.

However, by proceeding according to described reflection approach we end up with the following stationary worldline action
\begin{align}
 \mathcal{W}_0 \simeq \frac{E_\mathrm{S}}{E} \begin{cases} 
      \pi & \gamma < \check \gamma \\
      2 \check x_{4} \sqrt{1 - \check x_{4}^2 } + 2 \mathrm{arcsin}(\check x_{4}) & \gamma \geq \check \gamma 
   \end{cases},
   \label{eq:W0}
\end{align}
where 
\eqnsplit{
\check x_{4} = \frac{ \check \gamma + \delta}{ \gamma + \delta},\quad
\check \gamma = \left( \ln(1/\epsilon) \right)^\frac{1}{4N + 2},\quad
\delta = - \frac{\check \gamma}{\alpha} \frac{\mathfrak{Z}}{\mathfrak{D}}.
\label{eq:W0-adds}
}
In order to compute the remaining quantities $\alpha,\mathfrak{Z}$ and $\mathfrak{D}$ in \eqref{eq:W0-adds}, we Taylor expand the associated transcendental function in $\xi < 1$, where $\delta \equiv \xi \check \gamma $, see Sec. 5.3 in \cite{Akal:2017ilh}, and truncate the resulting series after the second order which leads to the following expressions
\begin{widetext}
\eqnsplit{
\mathfrak{D}
&:= 
2 \epsilon (2 N+1)  (2 \alpha  \Omega_2 +4 \alpha  N \Omega_2 +4 N \Omega_1 + 3 \Omega_1),\\
\mathfrak{Z}
&:=
2 \alpha  \Omega_1 \epsilon +4 \alpha  N \Omega_1 \epsilon +4 N+\Omega  \epsilon +2
+\big[
(\epsilon  (2 \alpha  \Omega_1+\Omega )+4 N (\alpha  \Omega_1 \epsilon +1)+2)^2\\
&-4 \alpha \epsilon (2 N+1) (4 N+\Omega  \epsilon +2) (2 \alpha  (2 N+1) \Omega_2 +(4 N+3) \Omega_1)
\big]^{1/2},\\
\Omega &:=  \mathbf{E}_{\frac{4 N+1}{4 N+2}  }(-\alpha),\quad
\Omega_1 :=  \mathbf{E}_{\frac{4 N+1}{4 N+2} - 1 }(-\alpha),\quad
\Omega_2 :=  \mathbf{E}_{\frac{4 N+1}{4 N+2} - 2 }(-\alpha),\quad
\alpha := {\check \gamma}^{4N+2}.
\label{eq:a-D-Z}
}
\end{widetext}
We begin with the correction $\delta$, which we expect to vanish for increasing $N$, here expressed as $N \uparrow$. The parameter $\xi$ is plotted versus $N$ in Fig.~\ref{fig:xi}, where the field strength ratio $\epsilon$ varies between different values as given in the plot legend. For $N=1$ the points clearly differ. But, as soon as $N \uparrow$, they rapidly merge together and converge to zero. Thus, the $\epsilon$ dependence becomes strongly suppressed and we find $\xi \rightarrow 0$, cf. Fig.~\ref{fig:xi}. Remarkably, such an $\epsilon$ independence applies usually for Sauter-like pulses which have a distinct pole structure in the instanton plane, cf. e.g. \cite{Linder:2015vta}. 

Super Gaussians do not share such properties, even for very large $N$, which is therefore an interesting coincidence in itself. We will come back to this point later on.
\begin{center}
\begin{figure}[h]
  \includegraphics[width=.49\textwidth]{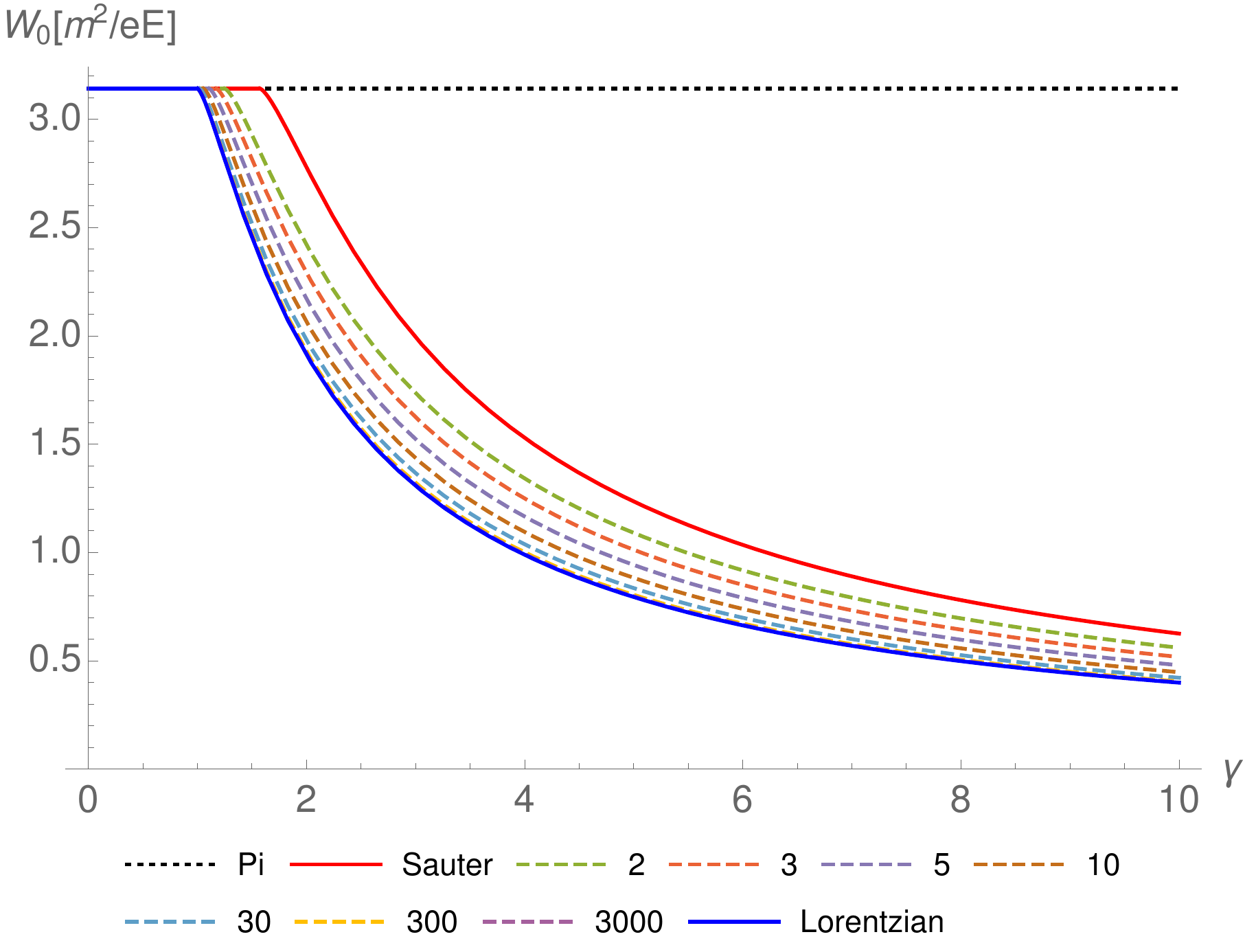}
\caption{Stationary worldline action $\mathcal{W}_0$ in units of $[E_\mathrm{S}/E]$. The integer values in the legend correspond to the parameter $N$ in \eqref{eq:f-g-functions}.}
\label{fig:W0}
\end{figure}
\end{center}
The nonperturbative prediction for the stationary worldline action in \eqref{eq:W0} is plotted in Fig.~\ref{fig:W0} versus $\gamma$, again for  different $N$ as listed in the plot legend, including the Sauter (red solid) and Lorentzian (blue solid) case. The dashed curves depict the predictions for the super Gaussian case.
Starting with $N=2$ (green), which already lies below the red solid curve, we find that as soon as $N \uparrow$ the curves converge to the blue solid one. For $N=3000$ (magenta) both results are visually indistinguishable. 
Furthermore, the critical threshold\footnote{The  critical  threshold  is  assumed to  be  determined  by  the  critical  point
where  both  the strong and the weak part start to contribute equally, see \cite{Akal:2017ilh}.}, which can be approximated\footnote{For this particular type of fields the $\Delta$ correction introduced in \cite{Akal:2017ilh} is negligible small, in particular for $N \gg 1$.} by $\check \gamma$ for large $N$ quite accurately, converges to $\gamma = 1$. Hence, for $N \rightarrow \infty$, corresponding to the usual rectangular potential barrier, we approach the blue solid curve as we have also seen in direct numerical computations\footnote{The accuracy of the analytical prediction in \eqref{eq:W0} increases as soon as $N \uparrow$. A similar behaviour applies for $\epsilon \downarrow$ with moderate $N$ as discussed in \cite{Akal:2017ilh} for $N \in  \{ 0,1 \}$.}. 
The numerically found threshold matches with our prediction $\check \gamma$.
We conclude that for parameters 
\begin{align}
N \in \mathbb{N}_{> 1}
\end{align}
the corresponding curves for $\mathcal{W}_0$ lie within the throat-like region bounded by the red (Sauter) and blue (Lorentzian) one, cf. Fig.~\ref{fig:W0}.

\section{Perturbative expansion}
\label{sec:perturb}
For weak Sauter-like pulses 
the first order contribution in perturbation theory respective $\epsilon$ turns out to be sufficient to reproduce the leading order exponential factor in $\mathcal{P}$.
If their Fourier transform in the large frequency limit
falls faster than exponentially, higher-order contributions become relevant. This observation has recently been made in \cite{Torgrimsson:2017pzs}.

In this context, we should note that finding out whether two different weak fields lead to the same tunneling exponent is not directly visible via their Fourier transforms, even in the large frequency limit.

Let us make this more concrete: for instance, both a Lorentzian and a Sauter pulse have transforms in the mentioned limit which decay exponentially, see expressions \eqref{eq:tild-g-approx-Lor} and \eqref{eq:tild-g-approx-Sauter} below. These functions are clearly distinguishable in form of a frequency shift by a factor $\frac{\pi}{2}$ and thus do not coincide.
So even for both the first order in the small parameter $\epsilon$ is sufficient to approach the nonperturbative result,
the corresponding stationary action $\mathcal W_0$ is distinct.

Now, according to the findings in \cite{Torgrimsson:2017pzs}, one may explain the aforementioned frequency shift and hence the impact on $\mathcal W_0$ via the approximate Fourier
transforms in the large frequency limit just by rescaling the estimate \eqref{eq:tild-g-approx-Lor}, i.e. $\omega \rightarrow  2\omega/\pi$, or vice versa. 
However, we want to emphasize that this case is rather special.

Namely, in Sec.~\ref{sec:nonperturb} we have shown hat the weak super Gaussian with $N \rightarrow \infty$ leads to the same stationary action as one obtains for the Lorentzian, cf. Fig.~\ref{fig:W0}.
So it is reasonable to expect that for such a rectangular pulse the first order in $\epsilon$ will be sufficient.
Along the lines of \cite{Torgrimsson:2017pzs}, such a behavior can already be anticipated, since the Fourier transform 
does not decay faster than an exponential. 

However, the obtained coincidence for the stationary action $\mathcal W_0$ cannot be unveiled just by working out the corresponding Fourier transforms which are indeed highly distinct, cf. \eqref{eq:fourier-lorentzian} and \eqref{eq:fourier-SG} as well as Fig.~\ref{fig:transforms}.
In contrast, as we have shown in Sec.~\ref{sec:nonperturb}, this result can be explained
by means of the corresponding effective reflection points in the instanton plane.

So, not only 
in order to support our results, but also to demonstrate in particular the differences occurring for any finite $N > 1$, which has not been analytically studied so far, we discuss in the following the super Gaussian from \eqref{eq:f-g-functions} in Fourier space.

\subsection{Fourier space}
Let $\widetilde{g}$ be the Fourier transform of the weak pulse. 
As mentioned, 
the order-by-order contributions in $\epsilon$ can be written in terms of $\widetilde{g}$.
For the Lorentzian we find
\eqnsplit{
\widetilde{g}(\varpi) &= 
\frac{1}{\omega} \sqrt{\frac{2}{\pi}}
\frac{\varpi}{\omega} \mathbf{K}_1\left( \frac{\varpi}{\omega} \right)
\label{eq:fourier-lorentzian}
}
with $\mathbf{K}_1$ being the first-order modified Bessel function of the second kind. 
For super Gaussians as in \eqref{eq:f-g-functions}, the representation in Fourier space is much more difficult to obtain.
Therefore, we need to introduce a slightly different strategy 
which may also be suitable for other 
backgrounds leading to similar problems.
We construct the super Gaussian ($\mathrm{SG}_{4N+2}$), particularly in the (almost) rectangular potential barrier limit, i.e. $N \gg 1$, which is the interesting case here, via the convolution of an ordinary Gaussian,
\begin{align}
\mathrm{G}_{\sigma_g} &\hat{=}\ e^{-( t/\sigma_g)^2 },
\end{align}
with the standard rectangular function,
\begin{align}
\mathrm{R}_{\sigma_r} \hat{=}\ \mathrm{rect}\left(\frac{t}{2 \sigma_r}\right).
\end{align}
So in order to compute $\widetilde{g}$, we proceed according to the following prescription
\begin{center}
\begin{tikzpicture}
  \matrix (m) [matrix of math nodes,row sep=3em,column sep=4em,minimum width=2em]
  {
     \mathrm{SG}_{4N+2} & \widetilde{\mathrm{SG}}_{4N+2} \\
     \frac{1}{{\mathrm C}_{\sigma_g,\sigma_r}} \left( \mathrm{G}_{\sigma_g} \otimes \mathrm{R}_{\sigma_r} \right) & \frac{1}{\widetilde{\mathrm C}_{\sigma_g,\sigma_r}} \left( \widetilde{\mathrm{G}}_{\sigma_g} \times \widetilde{\mathrm{R}}_{\sigma_r} \right) \\};
  \path[-stealth]
    (m-1-1) 
    edge [dashed] node [above] {$\mathcal{FT}$} (m-1-2)
    edge [-] node [left] {$\simeq$} (m-2-1)
    (m-2-1) 
    edge node [below] {$\mathcal{FT}$} (m-2-2)
    (m-2-2) 
    edge [-] node [right] {$\simeq$} (m-1-2);
\end{tikzpicture}
\end{center}
where $\otimes$ denotes the convolution product and $\mathrm C_{\sigma_g,\sigma_r},\widetilde{\mathrm C}_{\sigma_g,\sigma_r}$ are some normalization factors.
Identifying 
\eqnsplit{
N &\leftrightarrow 1/\kappa,\\
\sigma_r &\leftrightarrow 1/\omega,
}
with $\kappa := \sigma_g/\sigma_r$,
we can finally write (including the prefactor)
\eqnsplit{
\widetilde{g}(\varpi) = 
\frac{1}{\omega} \sqrt{\frac{2}{\pi}}
\frac{\omega}{\varpi} \sin\left(\frac{\varpi}{\omega}\right) \exp\left(-\frac{\kappa^2 \varpi^2}{4 \omega^2}\right)
\label{eq:fourier-SG}
}
having assumed the condition $\kappa \ll 1$. It is important to keep the parameter $\kappa$ for later purpose.

\subsection{First order in $\epsilon$}
The general expression after perturbing the interaction Hamiltonian in the Furry picture gives \cite{Torgrimsson:2017pzs}
\eqnsplit{
\mathcal{P} = V_3 \int \frac{dp^3}{(2 \pi)^3} \left| \ldots + \epsilon  \int \frac{d\varpi}{2\pi}\ \widetilde{g}\ \Pi_{\pmb{p}} + \ldots \right|^2.
\label{eq:P-perturb}
}
For simplifications we assume $\pmb{p} = 0$ which is reasonable, since the spectrum for backgrounds considered here is symmetrically peaked around the origin. Then the matrix element at $\mathcal{O}(\epsilon)$ takes the form 
\begin{align}
\Pi_{0}(\varpi) =
e^{ \frac{E_\mathrm{S}}{E} \left(  \left[ \frac{\varpi}{2 m} \sqrt{ 1 - \left( \frac{\varpi}{2 m} \right)^2 } + \mathrm{arcsin}\left( \frac{\varpi}{2 m} \right) \right] - \frac{\pi}{2} \right) }
\label{eq:Pi0}
\end{align}
which, not surprisingly, becomes unsuppressed for $\varpi = 2m$.
We begin with the Lorentzian pulse, $g(t) = \left[ 1 + (\omega t)^2\right]^{-3/2}$.
In order to perform a saddle point approximation to the $\varpi$ integral in \eqref{eq:P-perturb}, we first assume $\varpi \gg \omega$ in order to estimate
\begin{align}
\widetilde{g} \simeq \exp\left(-  \frac{\varpi}{\omega} \right).
\label{eq:tild-g-approx-Lor}
\end{align}
We insert the approximate expression \eqref{eq:tild-g-approx-Lor} into 
\eqref{eq:P-perturb} and find
the corresponding saddle point \cite{Torgrimsson:2017pzs}
\begin{align}
\varpi_\mathrm{sp} = 2 m \sqrt{1 - 1/\gamma^2}.
\label{eq:varpi_sp}
\end{align}
The latter leads to the previously introduced threshold $\gamma \geq 1$.
For $\gamma = 1$ the contribution $\widetilde{g}(\varpi_\mathrm{sp})$ is maximal where the exponential $\Pi_0(\varpi_\mathrm{sp})$ approaches its minimum. 
Defining a variable 
\eqn{
x := \varpi/\omega,
\label{eq:x}
}
we find the following solution for \eqref{eq:fourier-lorentzian}
\begin{align}
\int d\varpi\ \widetilde g = 
\sqrt{\frac{2}{\pi}}
\int_0^\infty dx\ x \mathbf{K}_1(x) &= 
\sqrt{\frac{\pi}{2}}.
\label{eq:K1-int}
\end{align}
\begin{figure}[h]
  \centering
  \includegraphics[width=.49\textwidth]{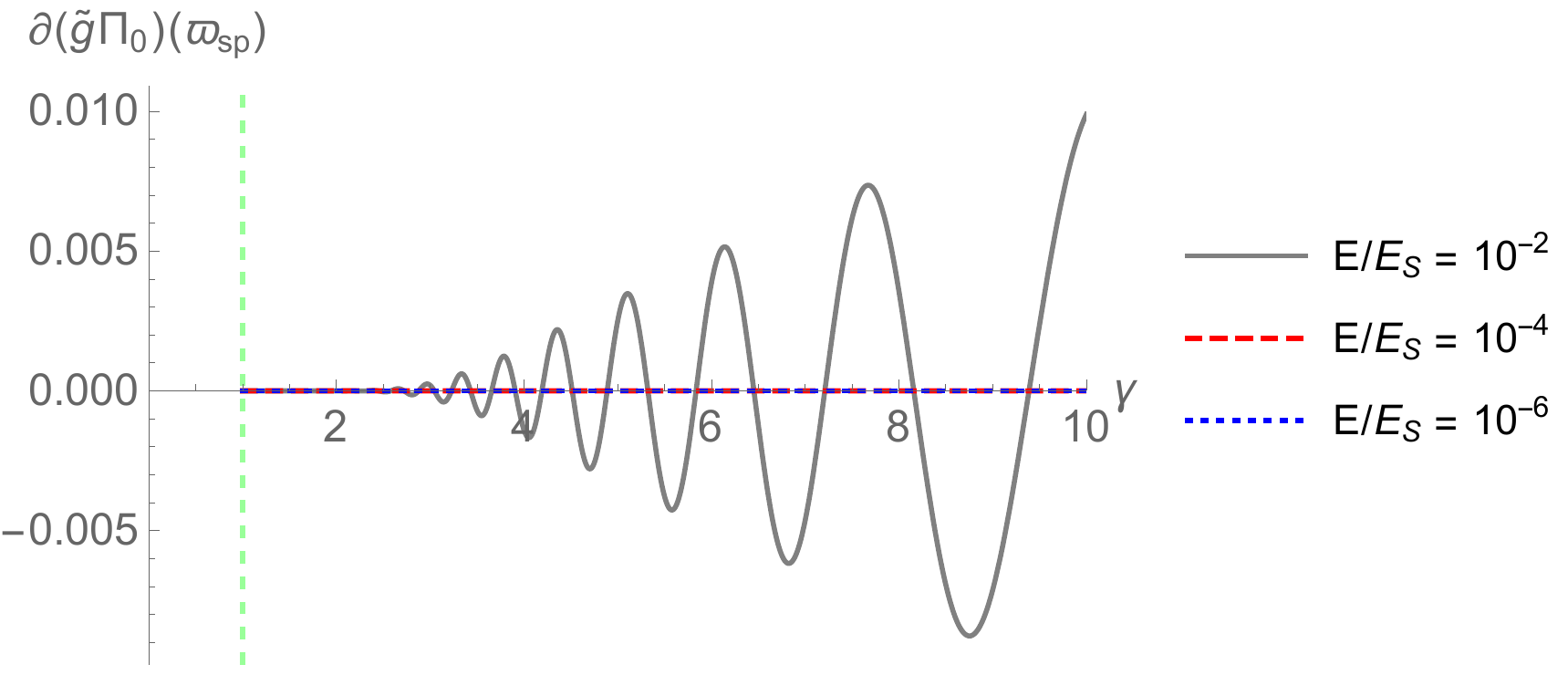}
\caption{Saddle point condition \eqref{eq:sp-cond} evaluated in $\varpi_\mathrm{sp}$ for different ratios $E/E_\mathrm{S}$ plotted versus $\gamma$. The vertical dashed line is placed at the critical threshold $\gamma = 1$.}
\label{fig:spts}
\end{figure}
In case of the super Gaussian we are particularly interested in the limit $\kappa \rightarrow 0$. 
For this we cannot write an exponential expression for $\widetilde{g}$ just by assuming $x \gg 1$.
However, according to the findings in Sec.~\ref{sec:nonperturb} we check whether $\varpi = \varpi_\mathrm{sp}$, see \eqref{eq:varpi_sp}, solves the saddle point condition 
\begin{align}
\partial (\widetilde g\ \Pi_0) \big|_{\kappa \rightarrow 0} = 0
\label{eq:sp-cond}
\end{align}
where $\partial \equiv \partial/\partial_\varpi$. 
It turns out that for the nonperturbative weak field regime, i.e. $E/E_\mathrm{S} \ll 1$ and $\omega \ll m$, the condition \eqref{eq:sp-cond} is fulfilled, cf. Fig.~\ref{fig:spts}.
For $E/E_\mathrm{S} = 10^{-2}$ and $\gamma \gtrsim 2$ the curve becomes increasingly oscillating until it settles down at $\simeq 0.15$.
Such a breakdown is reasonable, since according to $2 E/E_\mathrm{S} = \omega/m$
the gray solid curve with $\omega/m> 2 \times 10^{-2}$ almost approaches the Compton scale.
An approximate validity condition for $\varpi_\mathrm{sp}$ can be therefore given as
\begin{align}
\gamma E/E_\mathrm{S} \lesssim 10^{-2}
\end{align}
which is obviously satisfied for $E/E_\mathrm{S} = 10^{-4}$ (red, dashed) and $E/E_\mathrm{S} = 10^{-6}$ (blue, dotted) depicted in Fig.~\ref{fig:spts}.
Now, applying again the previous variable substitution to \eqref{eq:fourier-SG}, we obtain the same integral solution as in the Lorentzian case, cf. Eq.~\eqref{eq:K1-int}, 
\begin{align}
\int d\varpi\ \widetilde g = 
\sqrt{\frac{2}{\pi}} \int_0^\infty dx\ \frac{\sin(x)}{x} e^{-\kappa^2 x^2/4} \overset{\kappa \rightarrow 0}{=} \sqrt{\frac{\pi}{2}}.
\label{eq:si-int}
\end{align}
For large $x$ the integrand oscillates around the function in Eq.~\eqref{eq:fourier-lorentzian},
but asymptotically converges to zero.
Therefore, since $\varpi_\mathrm{sp}$ works for any $\omega$, at least for $\omega \ll m$, we may conclude that the threshold at $\gamma = 1$ applies for the super Gaussian in the limit $N \rightarrow \infty$ as well. This is exactly what we have found using the previous nonperturbative approach, see Sec.~\ref{sec:nonperturb},
which has also been confirmed in direct numerical computations.
Note that, as soon as $\kappa$ is taken to be sufficiently large, which basically corresponds to super Gaussians
with finite order $N$,
the latter coincidence will not apply anymore.
For completeness, let us discuss the Sauter pulse, $g(t) = \mathrm{sech}^2(\omega t)$, which has the following transform 
\eqnsplit{
\widetilde{g}(\varpi) =  
\frac{1}{\omega} \sqrt{\frac{\pi}{2}}
\frac{\varpi}{\omega} \text{csch}\left(\frac{\pi}{2 } \frac{\varpi}{\omega}\right).
\label{eq:fourier-sauter}
}
Again we can write an approximate expression assuming $x \gg 1$, 
\eqnsplit{
\widetilde{g} \simeq \exp\left( - \frac{\pi}{2} \frac{\varpi}{\omega} \right).
\label{eq:tild-g-approx-Sauter}
}
Inserting \eqref{eq:tild-g-approx-Sauter} and \eqref{eq:Pi0} into \eqref{eq:P-perturb}, results in the known critical threshold $\gamma \geq \pi/2$. Integrating the transform \eqref{eq:fourier-sauter} as before, we find
\eqnsplit{
\int d\varpi\ \widetilde g = 
\sqrt{\frac{\pi}{2}} \int_0^\infty dx\ x \mathrm{csch}\left( \frac{\pi}{2} x \right) = \sqrt{\frac{\pi}{2}}.
\label{eq:csch-int}
}
So the solution is identical to the previous one obtained for the Lorentzian and rectangular pulse in \eqref{eq:K1-int} and \eqref{eq:si-int}, respectively.

\subsection{Integral coincidence}
\begin{figure}[h]
  \centering
  \includegraphics[width=.4\textwidth]{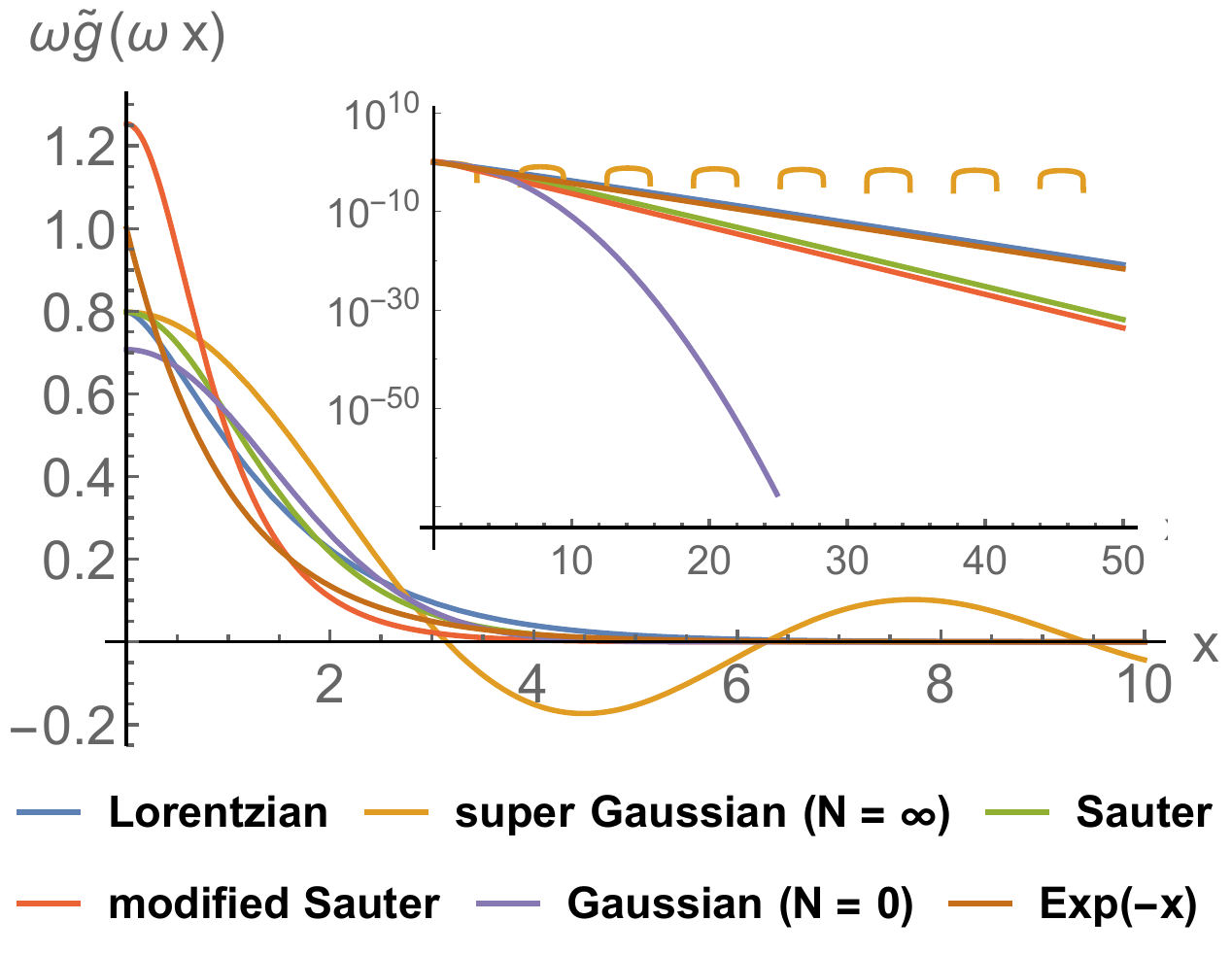}
\caption{(Rescaled) Fourier transforms $\omega \widetilde g(\omega x)$ where $x:= \varpi/\omega$ are plotted for the cases
indicated in the plot legend. 
For the ordinary Gaussian ($N=0$) we have $\sqrt{2} \omega \widetilde g(\omega x) = e^{-x^2/4}$. In the inset the same curves are shown with logarithmic scaling.}
\label{fig:transforms}
\end{figure}
Evaluating first the exponential via a saddle point approximation using
a large frequency estimation for $\widetilde g$,
we consider the integral $\int d\varpi\ \widetilde g$ as a prefactor in front of the leading order exponential \eqref{eq:Pi0} in \eqref{eq:P-perturb}.
Our findings above suggest that this integral 
seems to incorporate useful information about 
the impact of the additional weak dependence.
Namely, 
we have seen that
\eqnsplit{
\int_0^\infty dx\ \omega \widetilde{g}(\omega x) = \sqrt{\frac{\pi}{2}}
\label{eq:usef-int}
}
applies for all weak pulses (i.e. Lorentzian, super Gaussian with $N \rightarrow \infty$ and Sauter)
which approach the nonperturbative result already at lowest order $\mathcal O(\epsilon)$.
Notably, a Sauter pulse with frequency shift $\omega \rightarrow \omega \frac{\pi}{2}$ 
behaves similarly.

We want to make clear, that
the Fourier transforms $\tilde g$ as well as the condition \eqref{eq:usef-int} does not carry sufficient information unveiling whether the stationary action \eqref{eq:Gamma}
matches for two different backgrounds. 
On the other hand, we have seen in Sec.~\ref{sec:nonperturb}
that even when two backgrounds crucially differ in Minkowski spacetime, cf. Fig.~\ref{fig:profiles}, they can result in the same tunnelling exponential.

These insights can be taken as a strong evidence that the enormous dynamical enhancement is mainly triggered by the nonperturbative (effective) reflection points in the instanton plane, which in contrast to $\widetilde g(\varpi)$ depicted in Fig.~\ref{fig:transforms}, do perfectly agree.

Coming back to the integral condition in \eqref{eq:usef-int}, let us
adduce
an additional example. We consider a weak pulse of modified Sauter type described by 
\eqnsplit{
g(t) = \mathrm{sech}(\omega t),\qquad
\widetilde g(\varpi) = 
\frac{1}{\omega} \sqrt{\frac{\pi}{2}}
\mathrm{sech}\left( \frac{\pi}{2} \frac{\varpi}{\omega} \right).
\label{eq:fourier-mSauter}
}
In this case, we will find the same stationary action $\mathcal W_0$ for sufficiently small field strengths, usually $\epsilon < 10^{-2}$, as for the ordinary Sauter pulse \eqref{eq:fourier-sauter} which is rooted in the same reflection point, cf. \cite{Akal:2017ilh}. Therefore, we expect a same behavior with respect to $\epsilon$. Indeed, computing the corresponding integral, the result obeys again the condition in \eqref{eq:usef-int},
\eqn{
\int d\varpi\ \widetilde g = \sqrt{\frac{\pi}{2}} \int_0^\infty dx\ \mathrm{sech}\left( \frac{\pi}{2} x \right)
= \sqrt{\frac{\pi}{2}},
\label{eq:mSauter-int}
}
as in the previous cases.
We close this part by noting that for the super Gaussian with any finite order $N$ we will have
\eqn{
\int_0^\infty dx\ \omega \widetilde{g}(\omega x) < \sqrt{\frac{\pi}{2}},
}
since $\kappa > 0$, see \eqref{eq:fourier-SG}.
For such pulses, as will be shown in the following, higher orders in $\epsilon$ generally become relevant.
It is an interesting coincidence that in such a situation the condition
\eqref{eq:usef-int} 
is not fulfilled anymore.

\subsection{Higher orders in $\epsilon$}
For higher order contributions we 
rely on the general expansion
\begin{align}
\mathcal{P} \simeq \mathcal{P}_0 + \epsilon \mathcal{P}_1 + \epsilon^2 \mathcal{P}_2 + \mathcal{O}(\epsilon^3).
\end{align}
The zeroth order term stems again only from the strong background dependence.
The functions $\mathcal{P}_\mathfrak{N}$ can be obtained on basis of the $\mathfrak{N}$ photon master formula in a static background, see e.g. \cite{Schubert:2001he}. Performing a saddle point approximation with respect to the proper and worldline time, see Eq. (5.5) in \cite{Torgrimsson:2017pzs}, 
the leading order contribution reads
\eqnsplit{
 &\mathcal{P}_\mathfrak{N} \simeq \int d\varpi_1\ \widetilde{g}(\varpi_1)  \ldots \int d\varpi_\mathfrak{N}\ \widetilde{g}(\varpi_\mathfrak{N})\\ 
 &\times \exp\left( \frac{2m^2}{E} \left[ \Sigma \sqrt{1 - \Sigma^2} + \mathrm{arcsin}(\Sigma)  - \frac{\pi}{2} \right] \right)
 \label{eq:P_N-1}
}
where $0 < \Sigma < 1$ is defined as
\begin{align}
\Sigma := \frac{1}{2m} \sum_{i=1}^J \varpi_i,
\label{eq:Sigma}
\end{align}
and
\begin{align}
\sum_{l \in \{ 1,\ldots,J,\ldots,\mathfrak{N} \}} \varpi_l = 2m\Sigma + \sum_{j=J+1}^\mathfrak{N} \varpi_j = 0
\label{eq:energy-conv}
\end{align}
applies due to energy conservation. Note that the exponential in \eqref{eq:P_N-1} is of the same form as in \eqref{eq:Pi0}.
Without loss of generality let us assume $2 m \Sigma \gg \omega$. 
So for the Lorentzian we use again the approximate expression \eqref{eq:tild-g-approx-Lor} 
and compute the $\varpi_l$ integrals via \eqref{eq:energy-conv}.
Carrying out a saddle point approximation with respect to $\Sigma$  
results in
\eqnsplit{
&\mathcal{P}_\mathfrak{N} \simeq \exp\left(- \frac{4 m^2}{E} \frac{\Sigma_\mathrm{sp}}{\gamma} \right)\\
&\times  \exp\left( \frac{2m^2}{E} \left[ \Sigma_\mathrm{sp} \sqrt{1 -\Sigma_\mathrm{sp}^2} + \mathrm{arcsin}(\Sigma_\mathrm{sp})  - \frac{\pi}{2} \right] \right)
\label{eq:PN-saddles}
}
where $\Sigma_\mathrm{sp} = \sqrt{1 - 1/\gamma^2}$.
For the super Gaussian in the rectangular potential barrier limit, i.e. $\kappa \rightarrow 0$, the situation is not much different.
First, we solve the $\varpi_l$ integrals using condition \eqref{eq:energy-conv}.
The prefactor in front of the exponential in \eqref{eq:P_N-1} takes the form
\eqnsplit{
\prod_i \frac{\omega}{\varpi_i} \sin\left( \frac{\varpi_i}{\omega} \right)
\prod_j \frac{\omega}{\varpi_j} \sin\left( \frac{\varpi_j}{\omega} \right)
\label{eq:prefacs}
}
with
\eqnsplit{
\varpi_i &= \frac{2 m \Sigma}{J-1},\quad i\ \in \{2,\ldots,J\},\\
\varpi_j &= \frac{- 2 m \Sigma}{\mathfrak{N} - J - 1},\quad j\ \in \{J+1,\ldots,\mathfrak{N}-1\}.
}
In case of $2 m \Sigma \gg \omega$, we may use again the approximate form in Eq.~\eqref{eq:tild-g-approx-Lor}, since in the relevant regime it leads to the correct leading order contribution as we have seen before, see Fig.~\ref{fig:spts}.
The prefactors \eqref{eq:prefacs} in \eqref{eq:P_N-1} reduce then to an exponential that yields the following expression
\eqnsplit{
&\mathcal{P}_\mathfrak{N} \simeq \exp\left(- \frac{4 m \Sigma}{\omega} \right)\\
&\times \exp\left( \frac{2m^2}{E} \left[ \Sigma \sqrt{1 -\Sigma^2} + \mathrm{arcsin}(\Sigma)  - \frac{\pi}{2} \right] \right).
}
Rescaling $2 m \Sigma \rightarrow \Sigma$ subsequently, the saddle point is simply given by
$\Sigma_\mathrm{sp} = \varpi_\mathrm{sp}/(2m )$. This is the same exponential factor as in \eqref{eq:PN-saddles} which remains unchanged for any $\mathfrak{N} \geq 1$.

We conclude that similar as in the Sauter-like cases, the first order contribution in $\epsilon$ will be sufficient to approach the nonperturbative result.
Of course, this is quite different from the ordinary Gaussian, i.e. $N=0$, which behaves nonperturbatively, since higher orders in $\epsilon$ turn out to be necessarily relevant as discussed in \cite{Torgrimsson:2017pzs}.

\section{Summary and conclusion}
In this paper we have studied the Schwinger mechanism in the presence of an additional, uniformly oriented super Gaussian of integer order $4N+2$.

In the first part, Sec.~\ref{sec:nonperturb},
we have treated the respective background nonperturbatively. For doing so, we have utilized the reflection approach 
within the worldline formalism introduced in \cite{Akal:2017ilh} and derived purely analytical expressions in terms of $N$.
Specifically, we have shown that 
for $N=2$ a stronger dynamical enhancement applies in comparison to a weak contribution of Sauter type.
Taking the limit $N \rightarrow \infty$, which corresponds to the usual rectangular potential barrier,
has resulted in the same leading order exponential factor as one finds for a weak bell shaped Lorentzian.
Although both setups are highly distinct in Minkowski spacetime, the found coincidence applies due to
identical effective reflection points in the Euclidean instanton plane. 
Our findings thus demonstrate that such reflection points turn out to be the main regulator in this dynamical mechanism.
In this context, it is particularly interesting that even though the super Gaussian pulse for $N > 2$ results in a stronger dynamical enhancement compared to a Sauter pulse, we have found that in the former case with finite $N$ we still necessarily need higher order contributions in the field strength ratio $\epsilon$. This observation basically reflects the nonperturbative nature of this process. However, this turned out not to be the case for the Sauter pulse which, instead, behaves perturbatively in $\epsilon$. 
Hence, even though for moderate parameters such as $N \in \{2,3\}$ the super Gaussian becomes very close to the rectangular potential barrier, there is no perturbative behavior as for the Sauter and Lorentzian pulse, respectively.

In the second part, Sec.~\ref{sec:perturb}, we have studied the impact of the weak super Gaussian in perturbation theory
and found that in the limit $N \rightarrow \infty$ it shares the same higher order behaviour as Sauter-like pulses as recently seen in \cite{Torgrimsson:2017pzs}. For doing so, we have proposed a new approach based on convolution techniques. The latter have  substantially simplified the problem at hand which, in the way it has been approached in earlier studies, seems to be easily realizable for certain type of backgrounds for which computational difficulties in the Fourier space do not arise.
Referring to our findings, we have argued that the leading order contribution in the field strength ratio $\epsilon$ already approaches the nonperturbative result, although a distinct pole structure, as one finds for the Sauter-like pulses, is not present.
In addition, we have seen that for any finite $N$
a proposed integral condition is not fulfilled. Our findings have shown that the latter integral condition may serve as an indication of the relevance of higher orders in $\epsilon$ and thus may explain the nonperturbative behavior.
Our results have clearly demonstrated that tunnelling in such complex backgrounds can lead to nontrivial physics.
The fact whether the weak pulse behaves perturbatively or nonperturbatively with respect to $\epsilon$ depends on its microscopic details determined by the parameter $N$.
  
\section{Acknowledgments}
I thank Gudrid Moortgat-Pick for a careful reading of the manuscript.
I acknowledge the support of the Collaborative Research Center SFB 676 of the DFG.

\bibliographystyle{elsarticle-num}
\bibliography{article_bib}

\end{document}